\documentclass{appolb}
\usepackage{graphicx}
\usepackage{cite}
\usepackage{amsmath,amssymb}

\begin{document}
\title{Shape analysis of HBT correlations at STAR\thanks{Presented at the XIV Workshop on Particle Correlations and Femtoscopy, \\June 3 - June 7, 2019, Dubna, Russian Federation}}
\author{D\'aniel Kincses (for the STAR Collaboration)
\address{E\"otv\"os Lor\'and University\\P\'azm\'any P\'eter s\'et\'any 1/A, 1117 Budapest, Hungary}}

\maketitle
%\vspace{-2em}
\pagestyle{plain}
%\linenumbers
\begin{abstract}
To study the nature of the quark-hadron phase transition, it is important to investigate the space-time structure of the hadron-emitting source in heavy-ion collisions. Measurements of HBT correlations have proven to be a powerful tool to gain information about the source. In these proceedings, we report the current status of the analysis of source parameters obtained from L\'evy fits to the measured one-dimensional two-pion correlation functions in Au+Au collisions at $\sqrt{s_{NN}}$ = 200 GeV.

\end{abstract}

\section{Introduction}

Quantum-statistical (also called Bose-Einstein or HBT) correlations of identical bosons are used to explore the properties of the hot and dense matter created in heavy-ion collisions \cite{Lisa:2005dd}. These correlations can provide information on the space-time geometry of the particle-emitting source in heavy-ion collisions. 

Description of the shape of correlation function requires the knowledge of the source function which can be tested. Recent studies at different experiments \cite{Adare:2017vig,Sirunyan:2017ies,Porfy:2018mvd} showed that to properly describe the shape of the measured quantum-statistical correlation functions it is necessary to go beyond the Gaussian approximation. One possibility is to use L\'evy-stable distributions. There could be multiple (competing) reasons behind the appearance of such sources, like anomalous diffusion \cite{Metzler:1999zz, Csanad:2007fr}, jet fragmentation \cite{Csorgo:2004sr} or the proximity of the critical endpoint \cite{Csorgo:2003uv}. The definition of the one-dimensional L\'evy-stable distribution is the following \cite{nolan}:

\begin{equation}
f(x; \alpha, \beta, R, \mu) =\frac{1}{2\pi} \int_{-\infty}^{\infty}\varphi(q; \alpha, \beta, R, \mu)e^{iqx}dq,
\label{eq:fx}
\end{equation}
where the characteristic function is defined as:

\begin{align}
\varphi(q; \alpha, \beta, R, \mu) &= \exp\left(iq\mu-|qR|^\alpha(1-i\beta\textnormal{sgn}(q)\Phi)\right), \\ \Phi &= \left\{\begin{array}{l}\tan(\frac{\pi\alpha}{2}), \alpha \neq 1\\-\frac{2}{\pi}\log|q|, \alpha = 1\end{array}\right.
\label{eq:phi}
\end{align}

The four main parameters are the index of stability, $\alpha$, the skewness parameter, $\beta$ (the distribution is symmetric if $\beta = 0$), the scale parameter, $R$, and the location parameter, $\mu$. The latter is also the median of the distribution, and in case of $\alpha > 1$ it equals to the mean as well. The most important property of this distribution is that it retains the same $\alpha$ and $\beta$ under convolution of random variables, and any moment greater than $\alpha$ is not defined. In case of $\alpha < 2$ the distribution exhibits a power-law behavior, while the $\alpha = 2$ case corresponds to the Gaussian distribution. If we assume that the source is a centered, spherically symmetric L\'evy distribution ($S(x) = f(x; \alpha, 0, R, 0)$) and neglect any final state interaction, the one-dimensional two-particle correlation function takes the following form: 

\begin{equation}
C(Q) \approx 1+\lambda\frac{|\widetilde{S}(Q)|^2}{|\widetilde{S}(0)|^2} = 1+\lambda\cdot\exp{\left(-(RQ)^\alpha\right)},
\label{eq:cf}
\end{equation}
where $\widetilde{S}$ denotes the Fourier transform of the source, $Q$ is the one-dimensional relative momentum variable, defined as the absolute value of the three-momentum difference in the longitudinal co-moving system (for details see Ref. \cite{Adare:2017vig}), and $\lambda$ is the strength of the correlation function. 

\section{Results and discussions}

In this analysis, we have used Au+Au data at $\sqrt{s_{NN}}$ = 200 GeV recorded by the STAR experiment. We measured one-dimensional two-pion HBT correlation functions for like-sign pairs. For the experimental construction of the correlation functions we used the event-mixing technique. We applied the necessary event-, track-, and pair-cuts, similar to those used in Ref.~\cite{Adamczyk:2014mxp}. To incorporate the effect of the final-state Coulomb interaction, we used the Bowler-Sinyukov procedure \cite{Lisa:2005dd}:

\begin{equation}
C^{Coul.}(Q) = 1-\lambda+\lambda\cdot K(Q;\alpha,R)\cdot(1+\exp{\left(-(RQ)^\alpha\right))}
\label{eq:coul}
\end{equation} 

For the Coulomb correction, $K(Q;\alpha,R)$, a parametrized formula from Ref. \cite{Csanad:2019cns} was used. Fits of the correlation functions were preformed using the ROOT Minuit2Minimizer \cite{root}. 

\begin{figure}[h!]
\centerline{\includegraphics[width=0.9\textwidth]{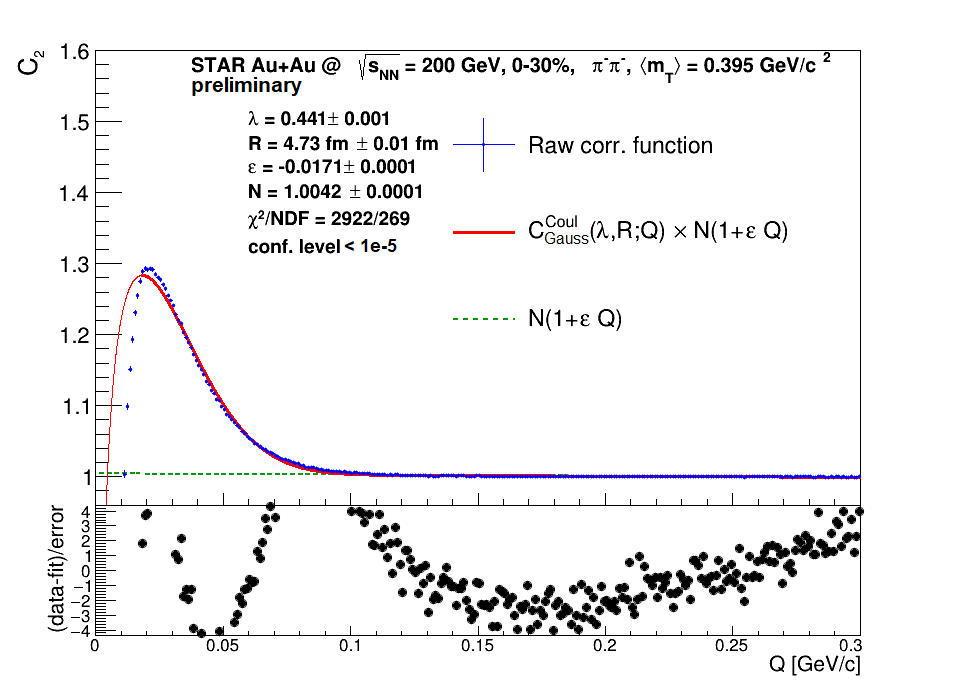}}
\caption{Example Gaussian fit (fixed $\alpha = 2$) of a Bose-Einstein correlation function of $\pi^-\pi^-$ pairs with a mean average transverse mass of $\langle m_T\rangle = 0.395$ GeV/c$^2$. The blue points correspond to the measured raw correlation function while the red curve is the fit function introduced in Eq.\ref{eq:coul}, complemented with a linear background.}
\label{f:gauss}
\end{figure}

\begin{figure}[h!]
\centerline{\includegraphics[width=0.9\textwidth]{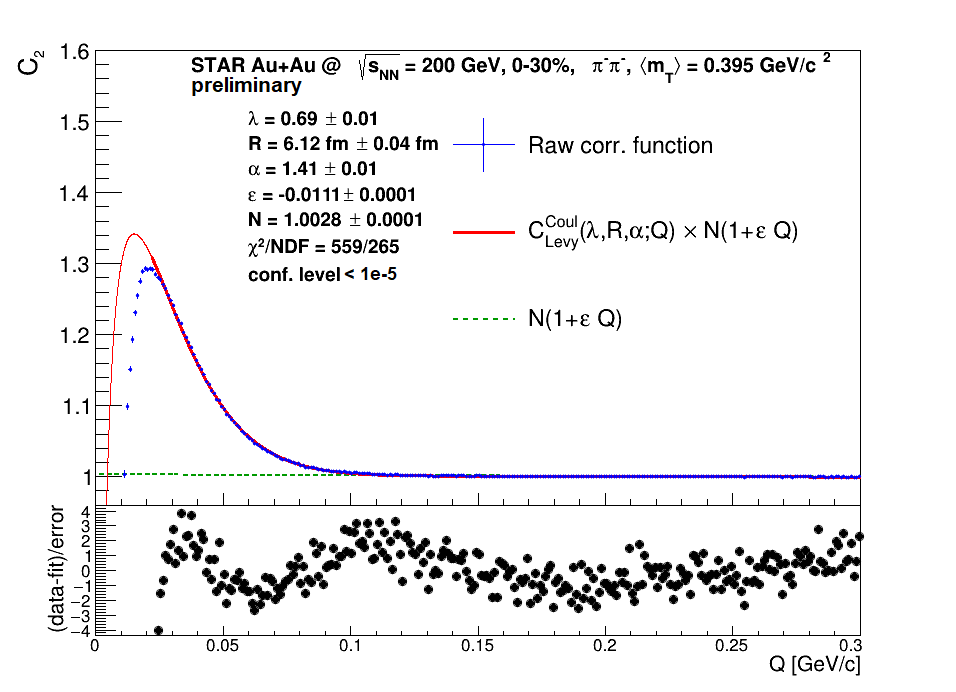}}
\caption{Example L\'evy fit of a Bose-Einstein correlation function of $\pi^-\pi^-$ pairs with a mean average transverse mass of $\langle m_T\rangle = 0.395$ GeV/c$^2$. The blue points correspond to the measured raw correlation function while the red curve is the fit function introduced in Eq.\ref{eq:coul}, complemented with a linear background.}
\label{f:levy}
\end{figure}

As a first check we investigated the Gaussian fits (with fixed $\alpha = 2$) to the data. The example is shown on Fig. \ref{f:gauss}. The value of $\chi^2$ is very high ($\chi^2/\rm{NDF}\sim 10$), the data are not described well by these fits. The magnitude of the L\'evy scale $R$ is compatible with the magnitude of the HBT radii extracted from three-dimensional Gaussian fits in Ref. \cite{Adamczyk:2014mxp}. Releasing the index of stability, $\alpha$, the $\chi^2$ values drop by a factor of 3-5, and the description highly improves in the $Q \gtrsim 25$ MeV/c region. Figure \ref{f:levy} shows the fit example with the $\alpha$ parameter released. An interesting observation is that the correlation function behavior at very low $Q$ is not described well by these kind of fits. Our investigations showed that this observation stands when varying the analysis cuts, and even in case of measuring the correlation function as a function of a different relative-momentum variable other than $Q$.

\newpage
\section{Summary and outlook} 

In these proceedings, we presented the first L\'evy-type HBT studies at STAR. We showed that indeed the Gaussian fits are not compatible with the measured data in case of one-dimensional two-particle correlation functions. The L\'evy fits provide a higher quality description of the data at $Q \gtrsim 25$ MeV/c, although the low $Q$ behavior is currently not clear. To understand the reason behind this observation, more detailed investigations are needed. These will include the detailed $m_T$ and centrality dependence, a thorough investigation of systematic uncertainties, and possibly the use of different expansion methods as suggested in Ref. \cite{DeKock:2012gp}. 

\section{Acknowledgments}

This research was supported by the \'UNKP-19-3 New National Excellence Program of the Hungarian Ministry for Innovation and Technology, as well as the NKFIH grant FK 123842.% and the funding agencies listed in Ref. \cite{funding}.


\newpage
\begin{thebibliography}{99}

%\bibitem{funding}
%STAR Funding Agencies: {\tt http://www.bnl.gov/rhic/STAR.asp}.

\bibitem{Lisa:2005dd}
M. A. Lisa {\it et~al.}, Ann. Rev. Nucl. Part. Sci. 55, 357 (2005)

\bibitem{Adare:2017vig}
PHENIX Collaboration, Phys. Rev. {\bf C97} no.6, 064911 (2018).

\bibitem{Sirunyan:2017ies}
CMS Collaboration, Phys. Rev. {\bf C97} no.6, 064912 (2018).

\bibitem{Porfy:2018mvd}
B. P\'orfy for the NA61/SHINE Coll., Acta Phys. Polon. B12, 451 (2018).

\bibitem{Metzler:1999zz}
R. Metzler, E. Barkai, and J. Klafter, Phys. Rev. Lett. 82, 3563 (1999).

\bibitem{Csanad:2007fr}%Levy, anomdiff
M. Csan\'ad, T. Cs\"org\H{o}, and M. Nagy, Braz. J. Phys. {\bf 37}, 1002 (2007).

\bibitem{Csorgo:2004sr}
T. Cs\"org\H{o} {\it et~al.}, Acta Phys. Polon. B36, 329 (2005).

\bibitem{Csorgo:2003uv}
T. Cs\"org\H{o} {\it et~al.}, Eur.Phys.J. {\bf C36}, 67 (2004).

\bibitem{nolan}
J. P. Nolan, Statistics and Probability Letters. 38, 187-195. (1998).

\bibitem{Adamczyk:2014mxp}
STAR Collaboration, Phys.Rev. {\bf C92} no.1, 014904 (2015).

\bibitem{Csanad:2019cns}
M. Csan\'ad, S. L\"ok\"os , M. I. Nagy, Universe 5, 133 (2019).

\bibitem{root}
F. James, M. Roos, Comput. Phys. Commun. 10, 343 (1975).

\bibitem{DeKock:2012gp}
M. B. De Kock, H. C. Eggers, T. Cs\"org\H{o}, PoS WPCF 2011 033 (2011).

\end{thebibliography}
\end{document}